\documentstyle[twocolumn,prl,aps,epsf]{revtex}
\def\beqn{\begin{eqnarray}}
\def\eeqn{\end{eqnarray}}
\def\beqns{\begin{eqnarray*}}
\def\eeqns{\end{eqnarray*}}
\def\beq{\begin{equation}}
\def\eeq{\end{equation}}
\def\bea{\begin{array}}
\def\ea{\end{array}}

\def\<{\langle}
\def\>{\rangle}

\begin{document}
\twocolumn[\hsize\textwidth\columnwidth\hsize\csname
@twocolumnfalse\endcsname
\draft \preprint{}
\title{Conduction mechanism and magnetotransport in multiwall carbon
nanotubes} \author{Stephan Roche${\ }^{\dagger}$, Fran\c cois 
Triozon${\
}^{*}$, Angel Rubio${\ }^{\ddagger}$ and
Didier Mayou${\ }^{*}$ }
\address{${\ }^{\dagger}$Commissariat \`a l'\'Energie
Atomique,
DRFMC/SPSMS,
Grenoble, France.\\
${\ }^{*}$ LEPES-CNRS, avenue des Martyrs BP166, 38042
Grenoble, 
France.\\
${\ }^{\ddagger}$ Departamento de F\'\i sica
Te\'orica,
Facultad de Ciencias-Universidad de Valladolid,
E47011 Valladolid, Spain.}
\maketitle
\begin{abstract}We report on a numerical study of
quantum diffusion over $\mu m$ lengths in defect-free
multiwall nanotubes. The intershell coupling allows the electron
spreading over several shells, and when their periodicities along the
nanotube axis are incommensurate, which is likely in real systems, the
electronic propagation is shown to be non ballistic. This results in
magnetotransport properties which are exceptional for a disorder free 
system, and which help to understand
the experiments.  
\end{abstract}
\pacs{PACS numbers: 73.50.-h, 73.61.Wp,72.80.Rj,
73.40.-c}
]


Carbon nanotubes are remarkable for their structure
and electronic 
properties, and seem suitable for molecular
electronic devices\cite{Ijima,Saito,CN-FET}. Single
wall
carbon
nanotubes (SWNTs) can be either metallic or
semiconducting depending on 
their helicity, i.e. how the graphene sheet is rolled
up\cite{Saito}.
Experimentally, metallic
SWNT are very good conductors, exhibiting ballistic 
transport\cite{CN-FET,Bachtold2}. Structural disorder
in
these systems is very small leading to mean free path
in the $\mu m$
range\cite{Todorov}.

Multiwall nanotubes (MWNTs) have a complex 
structure that consists of several shells sharing the
same axis. Their 
electronic
properties are complicated and less
understood\cite{Ebessen}. Indications
for ballistic
transport exist in some MWNTs\cite{MWNT-QQ}, but the
interpretation 
of quantized conductance $n(2e^{2}/h)$ with half integer 
values of n is controversial. Some authors propose an
effect of
contact between electrode and the nanotube\cite{Louie}
whereas others 
show that
transport through several shells is essential\cite{LandFQ}.
Many evidences for diffusive regime 
and quantum interferences effects are 
also found in other
experiments\cite{Bachtold2,langer96,AB-NT}. This has 
been 
interpreted as a result of scattering by structural or
chemical
disorder, 
but the mean free path has to be several orders of
magnitude smaller 
than in SWNTs. Even in the
limit of vanishingly small disorder, this raises
fundamental questions 
such as~: 
how many shells participate to 
transport, is it just the outer shell that carries
current or are inner shells important ? 
Is electronic transport ballistic in these systems ?

In this Letter, we report the first theoretical study
of quantum
diffusion on the $\mu$-meter scale for defect-free
MWNTs. The
effect of the interlayer 
coupling
on the propagation of the wavepackets is analyzed. As
the different 
shells periodicities along the nanotube axis are in
general incommensurate, 
the system probed by the electron is not periodic,
and we show that this 
yields intrinsic non ballistic transport. The
consequences of
this anomalous 
propagation on the magnetoconductance, are
investigated for a magnetic 
field parallel to the nanotube axis.
In particular, a
negative magnetoresistance at low field is found, as
well as 
oscillations of the conductance which are periodic
with the magnetic 
flux 
$\Phi$ through each tube with a periodicity
$\Phi_{0}/2$ ($\Phi_{0}$ 
the quantum flux). 
This offers an alternative explanation of the
experimental results of 
Bachtold et al. \cite{AB-NT} without assuming strong
disorder.


Our model hamiltonian is a tight-binding one
which is believed to provide a good description of the
electronic structure
of MWNTs. In 
this model, one
$p_{\perp}$-orbital per carbon atom is kept, with zero
onsite
energies, whereas constant 
nearest-neighbor hopping on each layer n (n.n.), and
hopping between 
neighboring 
layers (n.l.) are defined by~\cite{Saito2}:

{\small $$
{\cal H} = \gamma_{0} \Biggl[\sum_{i,j \hbox{ } n.n.}
|p_{\perp}^{j}\> \<
p_{\perp}^{i}| \Biggr] - \beta \Biggl[\sum_{i,j \in n.l.}
\cos(\theta_{ij})
e^{\frac{d_{ij}-a}{\delta}} |p_{\perp}^{j}\> \<
p_{\perp}^{i}|\Biggr]
$$}

\noindent
where $\theta_{ij}$ is the angle between the $p_{\perp}^{i}$
and $p_{\perp}^{j}$
orbitals, and $d_{ij}$ denotes their relative
distance. The
parameters used here are~: $\gamma_{0}=2.9 eV$, $a=3.34\AA$, 
$\delta= 0.45\AA$~\cite{Saito2}.
In this tight-binding approach, the differences
between SWNTs and MWNTs 
stem from the parameter $\beta$, as the limit
$\beta=0$ corresponds to 
uncoupled shells. Estimate gives
$\beta\simeq\gamma_{0}/8$~\cite{Saito2}, but in order to get insight in 
the
effect of $\beta$ on transport properties, the cases
$0\leq\beta\leq\gamma_{0}$ have been considered. Synthetized MWNTs 
contain
typically a few tenth of inner layers
but due to computer limitations, we have restricted our study to 2
and 3-wall nanotubes, taking the intershell distance of
$3.4\AA$ as in graphite. 

By means of a powerful O(N)
method based on a development of the operator 
$\exp(-iHt/\hbar)$ in a basis of orthogonal
polynomials~\cite{TriozonRM}, the
time-dependent Schr\"odinger equation for the evolution of wavepackets
(WP) $|\psi \>$ up to large time, is solved.  This allows us to
calculate the spreading of the WP, defined as
{\small $L_{\psi}(t) = {\small \sqrt{\< \psi|
(\hat{X}(t)-\hat{X}(0))^{2} |\psi 
\> }}$} over micron length scales
($\hat{X} (t)$ is the position operator along the
tube axis in the Heisenberg representation). We also
define the
time-dependent diffusion coefficients, by $D_{\psi}(t)
= 
L_{\psi}(t)^{2}/t$.
$D_{\psi}(\tau_{\phi})$ is the diffusivity along the
nanotube axis,
if at $\tau_{\phi}$ the electronic wavefunction
looses its phase memory due to some inelastic
scattering.
The diffusion coefficient at $\tau_{\phi}$ is also
connected to the Kubo conductivity $\sigma=(e^{2}/h)
\rho \<D(\tau_{\phi})\>$, where $\rho$ is the density
of states, and
$\<D(\tau_{\phi})\>$ the average of diffusion
coefficients for WP close 
to the Fermi level\cite{aronov}. The wavepackets $|\psi \>$ are chosen
localized on single sites of the nanotube at initial time, and by
averaging over 
many
sites, we obtain energy-averaged transport properties.
The average 
spreading $L(t)$ and the
average diffusion coefficient $D(t)$ are defined by
{\small $L(t)= {\small \sqrt{\< L_{\psi}^{2}(t)\> } }
=\sqrt{t D(t)}$}, where 
$\< \>$ denotes an average
over many wavepackets. This provides a good qualitative
picture of wavepacket propagation in MWNTs
constituted of conducting shells. Note that
experimental results 
suggest that
Fermi energies away from the charge neutrality point
are relevant\cite{collins,kruger}. 
Recently, Kr\"uger et al. obtained variations of
the Fermi energy of 
$\pm 1$ eV, 
corresponding to $10-15$ conducting channels instead
of $2$ per metallic layer 
\cite{kruger}.

An essential geometrical remark is that depending on
their helicities {\small $(n,m)$}, two shells of a
given MWNT are
commensurate (resp. incommensurate), if the ratio
between their 
respective unit cell lengths $T_{\small{(n,m)}}$ along
the tube axis is a rational 
(resp. irrational) number\cite{Saito}. From
geometrical considerations, one expects that
incommensurate structures are more likely than
commensurate structures.  As shown in this work, 
this issue has deep consequences on the conduction mechanism intrinsic 
to
MWNTs.


\begin{figure}[htbp]
\epsfxsize=7cm
\centerline{\epsffile{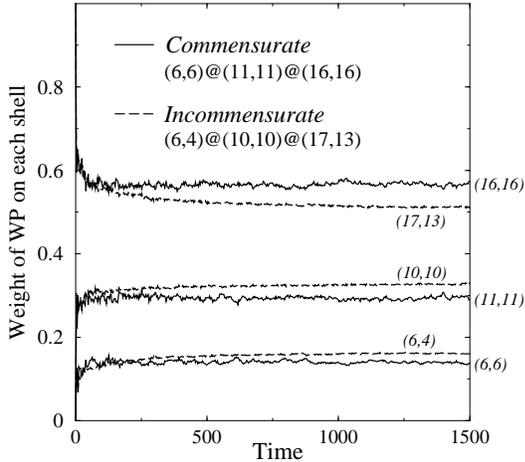}}
\caption{Temporal repartition of the wavepacket over the
3 shells. The initial state is localized on the outer shell
($(16,16)$ or $(17,13)$). Time is in
$\hbar /\gamma_{0}$ units and $\beta=\gamma_{0}/3$.}
\label{DIFF3WALL}
\end{figure}

As representative cases of 
commensurate systems, 
we have considered {\small $(9,0)@(18,0)$} and {\small 
$(6,6)@(11,11)@(16,16)$} 
($T_{\small{(9,0)}} = T_{\small{(18,0)}} = 3a_{cc}$
and 
$T_{\small{(6,6)}} = 
T_{\small{(11,11)}} = T_{\small{(16,16)}} =
\sqrt{3}a_{\small cc}$, 
where $a_{\small cc}
=1.42\AA$ is the interatomic distance between carbon
atoms).  
As representative cases of incommensurate systems, we
have considered
{\small $(9,0)@(10,10)$} and {\small
$(6,4)@(10,10)@(17,13)$} ($T_{\small{(6,4)}} =
3\sqrt{19}a_{\small 
cc}$, $T_{\small{(10,10)}} = \sqrt{3}a_{\small cc}$,
and
$T_{\small{(17,13)}}=3\sqrt{679}a_{\small cc}$).

{\it Interlayer electronic transfer}.- The spreading of
a WP, initially localized at one site of the outer shell, is
followed on
the different shells (Fig.\ref{DIFF3WALL}). Only two representative 
cases of 3-wall systems are reported. In commensurate systems, 
a rapid change in the weight of the wavefunction on each shell is
followed by a relaxation at the time scale of $\tau_{ll}\sim \hbar
\gamma_{0}/\beta^{2}$,
in good agreement with the Fermi Golden Rule. By
changing the amplitude 
of $\beta$ in the range $[\gamma_{0}/8 ,\gamma_{0}]$,
the expected scaling form of $\tau_{ll}$ is checked.
Note that for 
two
electrodes separated by $1\mu$m and assuming ballistic
transport with a 
Fermi velocity of $10^{6}ms^{-1}$, the corresponding time is $t \sim 
4500\hbar/\gamma_{0}$. This is two orders of magnitude
larger than 
$\tau_{ll}$ (for $\beta = \gamma_{0}/8$), and points
towards an important
contribution of interwall coupling in 
experiments. In the incommensurate systems, 
a continuous decay from the outer shells to the inner
shells is also found with similar characteristics.

\begin{figure}[htbp]
\epsfxsize=7cm
\centerline{\epsffile{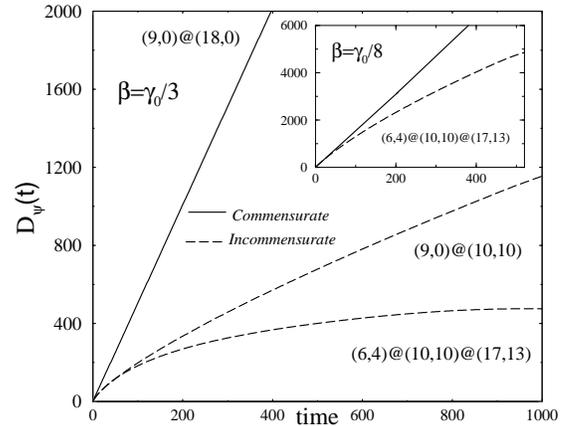}}
\caption{Main frame: Diffusion coefficients for typical 2-wall and
3-wall commensurate and
incommensurate MWNTs (for $\beta=\gamma_{0}/3$) as a function of  propagation
time (in $\hbar/\gamma_{0}$ units). Inset : Same
quantity for the $(6,4)@(10,10)@(17,13)$ with $\beta=\gamma_{0}/8$.}
\label{fig1} \end{figure}


{\it Quantum diffusion along the nanotube axis}.-A
fundamental difference between commensurate and incommensurate systems 
is
manifested in the quantum diffusion properties (Fig.\ref{fig1}). In
commensurate systems, $L(t)$ is found to follow a ballistic law~: $L(t) 
= vt
\rightarrow D(t) = v^{2}t$. This is expected 
on the basis of 
band structure theory for
periodic systems. The velocity $v \simeq 5.10^{5}
m.s^{-1}$, deduced numerically, is close to 
the averaged Fermi velocity of one isolated shell. It
is nearly independent on $\beta$ and depends weakly of
the
shells constituting the MWNT.

In the 2-wall incommensurate system, we observe 
an anomalous diffusion $L(t) 
\sim t^{\eta}$, $\eta=0.88$ (Fig. \ref{fig1}), for
$\beta= 
\gamma_{0}/3$, intermediate between
ballistic ($\eta=1$) and diffusive ($\eta=1/2$)
motions. 
Anomalous diffusion laws
are also observed in quasiperiodic systems, which
present a particular
repetitivity of local environnements
\cite{TriozonRM,QP}. 

In the 3-wall incommensurate systems, a saturation of
the diffusion coefficient is seen at large time, which
is equivalent to 
a
diffusive-like behavior ($L(t)\sim \sqrt{t}$) as
observed in disordered
systems. This allows to define an effective mean free
path $\tilde{l}_{e}$ and an effective scattering time
$\tilde{\tau}_{e}$ from $ \lim_{t\to
\infty}D(t)= D_{0}=\tilde{l}_{e}v =
v^{2}\tilde{\tau}_{e}$, where $v^{2}$ 
is the slope of $D(t)$ at origin. One can estimates an
effective elastic mean free path $\tilde{l}_{e}\simeq
35$ nm for $\beta = \gamma_{0}/3$ (Fig.\ref{fig1}). Note that a
diffusive law is also observed in the Harper model with a weak
incommensurate potential\cite{Harper}.

\begin{figure}[htbp]
\epsfxsize=7cm
\centerline{\epsffile{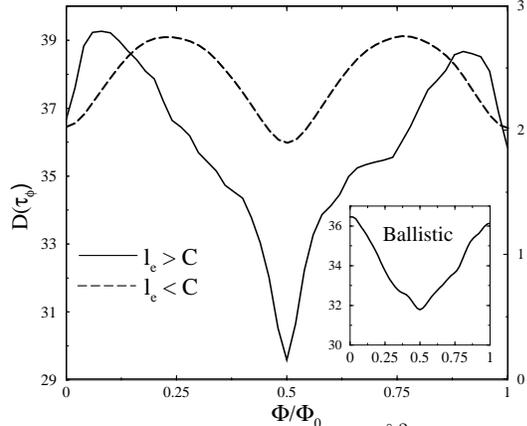}}
\caption{{\bf Main frame} :
$D(\tau_{\phi},\Phi)$
(in $\AA^{2}\gamma_{0}/\hbar$ unit) for a metallic
SWNT $(9,0)$ evaluated at time
$\tau_{\phi} \gg \tau_{e}$, for two disorder
strengths,
$V_{d}/\gamma_{0} = 3$ and $1$, such that
the mean free path ($l_{e}\sim 0.5$ and $3$ nm,
respectively) is either 
shorter (dashed
line) or larger (solid
line) than the nanotube circumference (${\cal C}\sim 2.3$nm).
The right
y-axis is associated to the dashed line and the left
y-axis to the 
solid
line. {\bf Inset}~: $D(\tau_{\phi},\Phi)$ for $l_{e}
= 3$ nm $>{\cal C}$
and $L(\tau_{\phi})<2l_{e}$.} 
\label{fig2}
\end{figure}

The striking difference 
between the 2-wall and 3-wall incommensurate cases, with the same 
hamiltonian parameters $\beta$ and $\gamma_{0}$, shows that the 
diffusion regime for
incommensurate MWNTs 
is very sensitive to the geometrical relation between
the different shells and to the number of shells. 
Contrary to the case of 
commensurate systems, in which the quantum diffusion
is weakly dependent on 
$\beta$, the diffusion law is sensitive to the
value of $\beta$. For 
example, for $\beta$ 
varying between $\gamma_{0}/3$ and $\gamma_{0}$, the diffusion exponent 
for
the 2-wall varies between $\eta = 
0.88$ and $\eta = 0.75$, and the estimated mean free
path for the 3-wall 
incommensurate system varies between $35 nm$ and $2
nm$. For
$\beta = 
\gamma_{0}/8$, the diffusive regime in the 3-wall is not reached in
the evolution time 
accessible to the computation (inset of Fig.\ref{fig1}).


{\it Transport in a magnetic field}.-Applying a
magnetic field
parallel to the MWNT axis, Bachtold et al.\cite{AB-NT}
have reported
negative magnetoresistance at low field (decrease of
the resistance with increasing
magnetic field strength) as well as $\Phi_{0}/2$
periodic
oscillations of the conductance, well described by the
weak localization (WL)
theory\cite{aronov}. They correlate this effect to an intrinsic 
disorder and
assume that the current is carried only by the outer nanotube shell.

In order to get insight in the
origin of these experimental results, we compare two
situations. 
In the first situation (case-{\bf I}), similar to the one considered by 
Bachtold et al.\cite{AB-NT},
the coupling between the outer and inner shells is
neglected. This corresponds to $\beta = 0$, which is equivalent to a 
SWNT.
A substitutional disorder is introduced by
a random modulation of onsite energies in the range
$[-V_{d}/2,V_{d}/2]$, in order to tune the value of 
the mean free path, since $l_{e} \propto 1/V_{d}^{2}$
in the limit
of weak scattering. In the second situation (case-{\bf II}), we 
consider a
MWNT with $\beta 
\neq 0$ but we 
neglect disorder. 

The magnetic field
modifies the hamiltonian through
a Peierls substitution~: hopping terms have a
field-dependent phase 
$e^{i\varphi_{ij}}$ with $\varphi_{ij} =
\frac{e}{h} \int_{i \rightarrow j}
\mathbf{A}$$.d\mathbf{r}$, where
$\mathbf{A}$ is  the vector potential.
We look at $D(\tau_{\phi},\Phi )$, where
$\Phi$ is the flux of the magnetic field through the tubes and is
identical for all tubes. It is sufficient to consider $0<\Phi<\Phi_{0}$ 
since at least a
$\Phi_{0}$-periodicity is expected from gauge invariance
\cite{Byers}. We assume that variations of the
diffusivity give the main contribution to
magnetoconductance.


\underline{Case-{\bf I}} (see Fig.3). Several types of
situations occur depending on the
relative values of $L(\tau_{\phi})$, $l_{e}$, and
$\cal{C}$, which are respectively the average spatial
extension of WP at the time $\tau_{\phi}$, the
elastic mean free path, and the circumference of the
nanotube. When $l_{e}<C<L(\tau_{\phi})$, our calculations
confirm that the behavior of $D(\tau_{\phi},\Phi )$
follows WL theory predictions. The diffusivity increases at
low field (negative magnetoresistance) and presents
a $\Phi_{0}/2$-periodic  Aharonov-Bohm oscillation,
i.e.
$D(\tau_{\phi},\Phi+\Phi_{0}/2)=D(\tau_{\phi},\Phi)$ 
(see
Fig.3). For a smaller
disorder with $l_{e} > C$, depending 
on the value of $\tau_{\phi}$, two different behaviors
are obtained. If $L(\tau_{\phi}) <
2l_{e}$,
then the diffusivity decreases at low field, and a $\Phi_{0}$-periodic
oscillation dominates the oscillative behavior of the
diffusivity, i.e.
$D(\tau_{\phi},\Phi+\Phi_{0})=D(\tau_{\phi},\Phi)$
(inset
Fig.3). This situation with large mean free path and
irrelevance of
backscattering indeed leads to a positive
magnetoresistance and
$\Phi_{0}$-periodic oscillation, in agreement with
theory\cite{Ando}. 
Whenever $L(\tau_{\phi}) > 2l_{e}$, 
although Aharonov-Bohm oscillations remain
$\Phi_{0}$-periodic, we get a 
negative
magnetoresistance at low magnetic field, which is
consistent with the 
fact
that $\tau_{\phi}$ is now large enough to allow some
backscattering to 
be
efficient. We also remark that the
marked reduction of diffusivity which shows up at $\Phi/\Phi_{0}=1/2$
is consistent with a study on small metallic cylinder\cite{Carini}. 

\underline{Case-{\bf II}} (see Fig.4). The case of the incommensurate 
2-wall $(9,0)@(10,10)$ is given in
Fig.4 (inset) 
where the power-law diffusion takes place. 
$\Phi_{0}$-periodic oscillations
and positive magnetoresistance at low field are
observed, similar to 
what is obtained for the ballistic regime in the SWNT
case. Conversely,
for $(6,4)@(10,10)@(17,13)$, which present a
diffusive-like 
propagation, 
there is evidence for negative magnetoresistance at
low field and 
$\Phi_{0}/2$-periodicity of
$D(\tau_{\phi},\Phi)$. The effective elastic mean free
path $\tilde{l}_{e}\simeq 35$ nm (see above), turns out to
be larger than the outer shell circumference ($\simeq
7nm$). Even in 
this situation, a small $\Phi_{0}/2$-periodic
oscillation is observed
for large
$\tau_{\phi}$. By taking an enhanced value of the
coupling parameter
($\beta=\gamma_{0}$), the diffusive regime is reached
more rapidly,
$\tilde{l}_{e} \sim 2nm < \cal{C}$, and together with
a 
negative
magnetoresistance at low field, a stronger
$\Phi_{0}/2$-periodic 
oscillation is obtained. These results confirm that magnetotransport 
properties
of MWNTs are very sensitive to the geometry, the number of shells 
carrying
current, and the hamiltonian parameters.

\begin{figure}[htbp]
\epsfxsize=7cm
\centerline{\epsffile{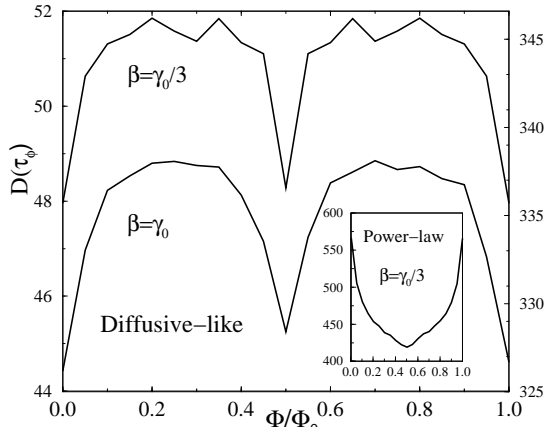}}
\caption{$D(\tau_{\phi},\Phi)$ (in
$\AA^{2}\gamma_{0}/\hbar$ unit) of the incommensurate 3-wall (main 
frame), at
time $\tau_{\phi}=3000\hbar
/\gamma_{0}$ for $\beta=\gamma_{0}/3$ (upper curve, right y-axis),
and at time
$\tau_{\phi}=1200\hbar /\gamma_{0}$ for
$\beta=\gamma_{0}$ (lower curve, left 
y-axis). Inset: Same quantity for the 2-wall at $\tau_{\phi}=1200\hbar.
/\gamma_{0}$}
\label{fig3}
\end{figure}

Note that 
in disordered systems, the basic
scheme is that of ballistic electrons scattering on random
impurities. The 
$\Phi_{0}/2$ periodicity results from
quantum interferences of the electronic pathways
around the cylinder 
wrist. This scheme is not strictly applicable in the
case of the 
incommensurate MWNTs, which
means that a 
clear theoretical understanding of the $\Phi_{0}/2$
periodicity has 
still to be 
developed, although it is obvious that the diffusive-like regime 
propagation
plays a central role.

{\it Conclusions}.-The above results demonstrate that the behavior
observed by Bachtold et 
al.\cite{AB-NT}
can be reproduced by considering only the effect of
incommensurability.  
In a real experiment with uniform magnetic field, the
flux through a 
given shell is proportionnal to its section, whereas
here we have taken 
identical flux for all shells. Our hypothesis is
relevant if, in the experimental conditions, the current is
confined on a few 
shells close to the outer one of the MWNT. This confinment could stem 
from the fact
electrons can not go through semiconducting shells, that represent 
statistically 
$2/3$ of the shells. Note that in the
interpretation 
developed by Bachtold et al., and according to the
above results (case-{\bf I}), the mean 
free path induced by disorder has to be smaller than
the circumference of 
the nanotube in order to recover the $\Phi_{0}/2$ periodicity. Yet,
using the formula of White and Todorov \cite{Todorov} with
the same disorder parameter, the mean free
path is four
orders of magnitude larger 
than the circumference
of the outer nanotube, in the 
experiment of Bachtold et al. \cite{AB-NT}.


In summary it has been shown that in MWNTs, the
electron wavepacket can spread over several 
shells, due to interlayer coupling. When the ratio
between the 
different
shells periodicities along the nanotube axis is
irrational,
the MWNTs are incommensurate systems. In such a situation, the 
electronic propagation is not ballistic but depends
very much on the geometry and number of shells participating to 
transport. 
It has been shown that incommensurability offers an alternative 
explanation
of the experimental results of Bachtold et al.\cite{AB-NT} on
magnetoresistance, without assuming strong disorder.

MWNTs could provide an unique opportunity of investigating transport in
incommensurate quasi one-dimensionnal conducting systems.

\noindent
Acknowledgments: Financial support from NAMITECH
[ERBFMRX-CT96-0067(DG12-MITH)], DGES (PB98-0345), JCyL
(VA28/99) and C$^4$ is acknowledged. S.R. is indebted
to R. Saito for stimulating encouragements.



\begin{thebibliography}{12345678}
\bibitem[\dagger]{add}To whom correspondence should be addressed
(sroche@cea.fr) 
\bibitem{Ijima}
S. Ijima, Nature {\bf 354}, 56 (1991).
\bibitem{Saito}
R.~Saito, G.~Dresselhaus, and M.~S. Dresselhaus, {\it
Physical Properties of Carbon Nanotubes} (Imperial
College Press,
London,
1998).
\bibitem{CN-FET}
S. Tans et al., Nature {\bf 393}, 49 (1998). R. Martel
et al., Appl.  Phys. Lett. {\bf 73}, 2447 (1998). 
\bibitem{Bachtold2}
A. Bachtold et al., Phys. Rev. Lett. {\bf 84}, 6082
(2000).
\bibitem{Todorov}
C.T. White and T. N. Todorov, Nature {\bf 393}, 240
(1998).
\bibitem{Ebessen}
T.W. Ebessen et al., Nature {\bf 382}, 54 (1996). 
\bibitem{MWNT-QQ}
S. Frank et al., Science {\bf 280}, 1744 (1998).
\bibitem{Louie}
H.J. Choi et al., Phys. Rev. B. {\bf 60} R14009 (1999).
\bibitem{LandFQ}
S. Sanvito et al., Phys. Rev. Lett. {\bf 84} 1974
(2000).
\bibitem{langer96}
L.~Langer et al., Phys. Rev. Lett. {\bf 76}, 479
(1996).
\bibitem{AB-NT}
A. Bachtold et al., Nature {\bf 397}, 673 (1999). 
\bibitem{Saito2}
R.~Saito, G.~Dresselhaus, and M.~S. Dresselhaus, J.
Appl. Phys. {\bf 73}, 494 (1993). J.C. Charlier and
J.P. Michenaud, Phys. Rev. Lett. {\bf 70}, 1858 (1993). 
Ph. Lambin, V. Meunier and A. Rubio, Phys. Rev. B {\bf 62}, 5129
(2000).
\bibitem{TriozonRM}
S. Roche and D. Mayou, Phys. Rev. Lett. {\bf 79}, 2518
(1997). S. Roche, Phys. Rev. B {\bf 59}, 2284 (1999).
F. Triozon et al., RIKEN Review {\bf 29}, 73 (2000).
\bibitem{aronov}
A.G. Aronov and Y.V. Sharvin, Rev. Mod. Phys {\bf 59},
755 (1987).
\bibitem{collins}
P.G. Collins et al., Science {\bf 287}, 1801 (2000).
\bibitem{kruger}
M. Kr\"uger et al., to appear in Appl. Phys. Lett.
\bibitem{QP}
B. Passaro et al., Phys. Rev. B {\bf
46}, 13751 (1992). 
\bibitem{Harper}
H. Hiramoto and M. Kohmoto, Phys. Rev. B {\bf 40},
8225 (1989).
\bibitem{Byers}
N. Byers and C.N. Yang, Phys. Rev. Lett. {\bf 7}, 46 (1961).
\bibitem{Ando}
T. Ando, Semicond. Sci. Technol. {\bf 15},  R13
(2000).
\bibitem{Carini}
J.P. Carini et al., Phys. Rev. Lett. {\bf 53}, 102
(1984).
\end{thebibliography}
\end{document}